\documentclass[]{MathAppl24}
\usepackage[OT4,T1]{fontenc}
\usepackage[utf8]{inputenc}
\usepackage{polski}
\usepackage{graphicx}

\usepackage{amsmath}
\usepackage{amssymb}
\usepackage{filecontents}

\usepackage[english,polish]{babel} 
\usepackage{lastpage} 

\usepackage{etex} 

\usepackage{color} 
\usepackage{cite} 
\usepackage{algorithm}
\usepackage{algorithmicx}
\usepackage{algpseudocode}
\usepackage{listings}
\usepackage{bbm}

\usepackage[table]{xcolor}
\usepackage{graphicx}
\usepackage{float}
\usepackage{multicol} 
\usepackage{multirow} 
\usepackage{makecell}
\usepackage{longtable} 
\usepackage{mathrsfs}
\usepackage{subfig}
\usepackage{array}
\newcommand\ChangeRT[1]{\noalign{\hrule height #1}}
\newcolumntype{M}[1]{>{\centering\arraybackslash}p{#1}}

\RequirePackage[numbers]{natbib}

\usepackage[colorlinks=true]{hyperref}
  \hypersetup{
    pdftitle={Identifying clusters in Czekanowski’s diagram}, 
    pdfauthor=Krzysztof Bartoszek, 
    colorlinks,
    urlcolor=blue,
    filecolor=magenta,
    citecolor=green, 
    linkbordercolor={1 1 1}, 
    citebordercolor={1 1 1},  
    urlbordercolor={ 1 1 1}  
  } 
 \RequirePackage[hyperpageref]{backref} 
    \renewcommand*{\backref}[1]{}  
    \renewcommand*{\backrefalt}[4]{
       \ifcase #1 
          No cited.
       \or
          Cited on p. #2.
       \else
          Cited on pp. #2.
       \fi}  
\newcommand{\orcid}[1]{\href{https://orcid.org/#1}{\includegraphics[scale=.05]{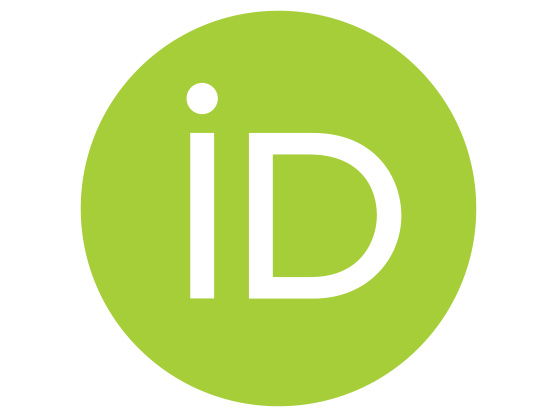}}}

\def\repo{http://wydawnictwa.ptm.org.pl/index.php/matematyka-stosowana/article/viewArticle}  

\pdfoutput=1
\usepackage{graphicx}
\usepackage{wrapfig}
\graphicspath{{./Figures/},{./Pictures/},{./Bio/}}
\usepackage{multicol}
\firstpage{i}    
\LogoG{\includegraphics[width=0.18\textwidth]{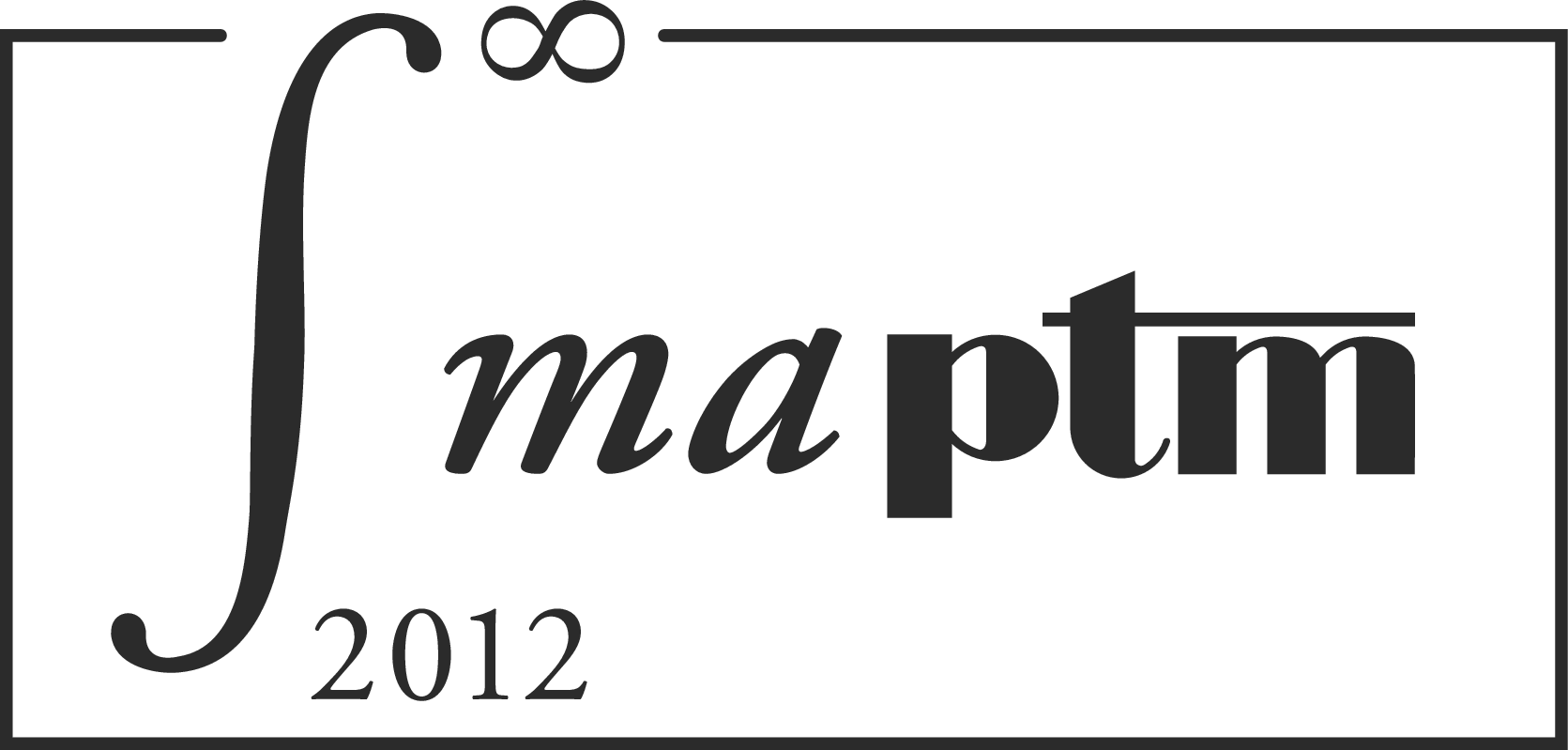}}
\volume{51}
\fasc{2}
\years{2023}
\LogoGMS{\includegraphics[width=0.18\textwidth]{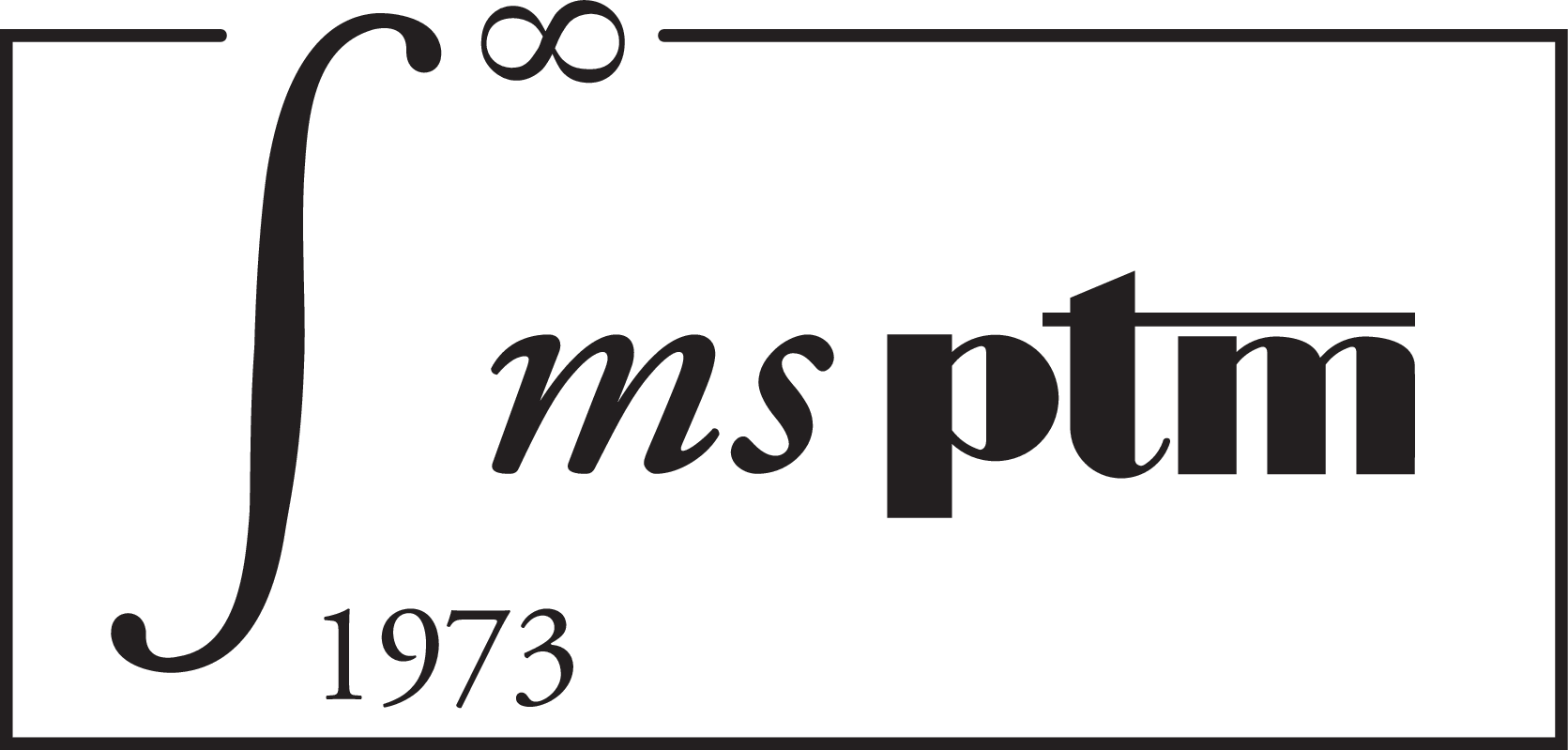}}
\volumeMS{35}
\numberMS{66}
\wwwtrue

\secnameMS{MATHEMATICAL BIOLOGY}
\pages{183--198}
\received[23rd of January 2023]{3rd of October 2023}
\lastrevision{}
\logo{}{}{}
\def\doinum{10.14708/ma.v51i2.7259}

\title[Clusters in Czekanowski’s diagram]{Identifying clusters in Czekanowski’s diagram}

\tpauthortrue 
\author[K. Bartoszek]{K.~Bartoszek\orcid{0000-0002-5816-4345}}
\affiliation{Link\"oping University}
\affiliation{Department of Computer and Information Science}
\address{The Division of Statistics and Machine Learning\\
\indent Link\"oping University, 581 83, Link\"oping, Sweden 
}
\city{Link\"oping}
\email{krzysztof.bartoszek@liu.se; krzbar@protonmail.ch}
\author[Y. Luo]{Ying Luo\orcid{0009-0002-1298-0037}}
\affiliation{Linköping University}
\address{The Division of Bioinformatics, Department of Physics, Chemistry and Biology\\
\indent 581 83, Linköping, Sweden}
\email{ying.luo@liu.se, kelly.yingluo@gmail.com}
\city{Link\"oping}

\comm{Urszula Foryś} 

\usepackage{algorithm}
\usepackage{algorithmicx}
\usepackage{algpseudocode}
\usepackage{listings}
\usepackage{bbm}

\newcommand{\pkg}[1]{\textbf{#1}}
\newcommand{\proglang}[1]{\textsf{#1}}
\newcommand{\code}[1]{\texttt{#1}}

\algnewcommand\algAnd{\textbf{and}~}
\algnewcommand\algOr{\textbf{or}~}

\subjclass[2010]{Primary: 62H30; Secondary: 62-04}
\keywords{Change point analysis, cluster analysis, Czekanowski’s diagram, fuzzy clustering, multivariate distance methods, \proglang{R}, \pkg{RMaCzek}.}

\begin{document}
\vspace{-5ex}
\setcounter{page}{183} 
\selectlanguage{english}\Polskifalse

\begin{abstract}
Visualizing data through Czekanowski's diagram has as its aim the illustration of the relationships between objects. Often, obvious clusters of observations are directly visible. However, it is not straightforward to precisely delineate these clusters. This paper presents the development of the package \pkg{RMaCzek}, which now includes features for cluster identification in Czekanowski diagrams.
\end{abstract}

\section{Introduction}
\label{sec:introduction}
\begin{figure}[H]
	\centering
    \includegraphics[width=0.745\textwidth]{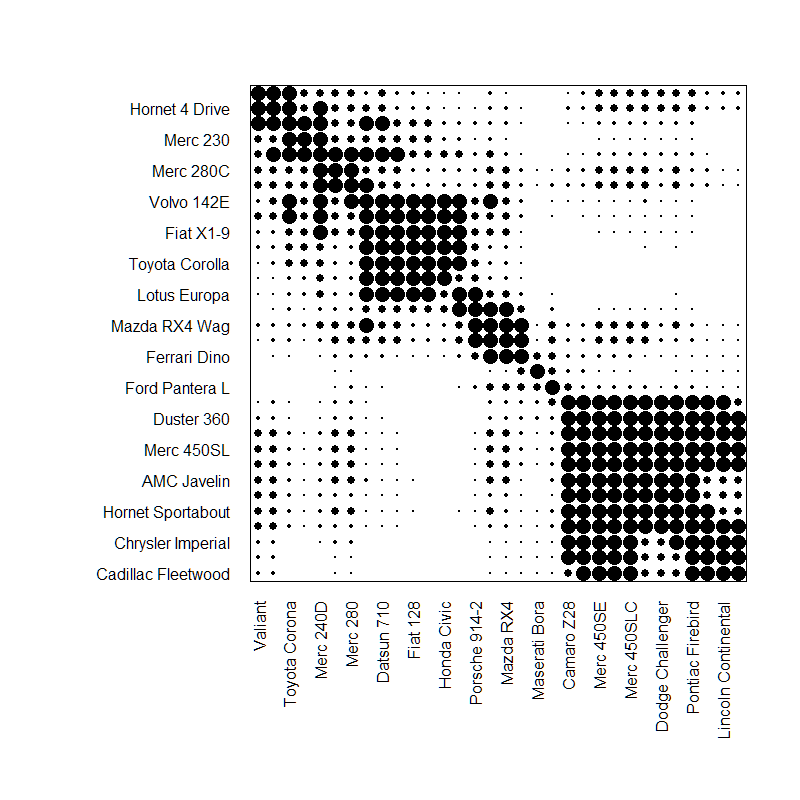} 
	\caption{
 Example symmetric Czekanowski's diagram based on the \code{mtcars} dataset \cite{HHenPVel1981} bundled with \proglang{R}. The clear bottom--right cluster are the larger horsepower cars. 
 }
	\label{fig:ExampleRMaCzek}
\end{figure}

The Polish anthropologist, Jan Czekanowski, introduced one of the earliest methods for taxonomic and proximity visualization methods in $1909$ \cite{JCze1909}. The result is a graphical representation of a matrix (refer to Figs.~\ref{fig:ExampleRMaCzek} or~\ref{fig:OLOWBCCzkDiag}), where the similarities and differences between observations are visually discernible.
The method first requires computing the distance matrix between observations in the dataset, then classifying the distance values into several groups, and finally plotting the matrix using symbols or colors. 
By rearranging the rows and columns of the distance matrix to bring similar observations closer and distant ones farther apart (a \emph{seriation} problem), the method enhances human visual perception, enabling a more effective interpretation of the data.

Statistical software that allows for the possibility of obtaining such diagrams has been implemented, most notably the \pkg{MaCzek} program \cite{ASolPJas1999}. Recently, a \proglang{R} \cite{R} for this purpose has been made available on CRAN, the \pkg{RMaCzek} package \cite{KBarAVas2020}.
This package has been used \cite{KBar2022KKZMBM} to replicate Jan Czekonowski's original skull classification study, which motivated his original introduction of the method \cite{JCze1909}. In our work here, we describe further developments of this \proglang{R} package \cite{YLuo2022}.

The creation of \emph{Czekanowski's diagram} has already been thoroughly described in the literature (e.g. \cite{KBarAVas2020, JCze1909, ASolPJas1999}, Ch, V in \cite{JCze1913}, and Ch. $2.2$ in \cite{YLuo2022}). 
Hence, here we will only briefly make the previous, informal, description more precise.
The process starts by obtaining a serialized matrix, $D\in \mathbb{R}^{N \times N}$, where $N$ is the number of observations. This is achieved by calculating the distance matrix between observations and rearranging the rows and columns based on the chosen serialization method.
In other words, similar samples are positioned closer in $D$, while dissimilar ones are placed further apart.
Then, the matrix $D'\in \mathbb{R}^{N \times N}$ is generated by discretizing the values of $D$, and assigning a symbol to each class (see Fig.~\ref{fig:ExampleRMaCzek}). This so--called \emph{Czekanowski's matrix} can be constructed symmetrically or asymmetrically. The asymmetric approach was originally proposed by Czekanowski in 1909 \cite{JCze1909}.
According to it, each column is treated separately and each row entry is assigned symbols depending on whether it is in the group of closest observations to the focal, column observation; next closest group; next, next closest group, e.t.c.
In the symmetric case, the distance values in $D$ would be divided into a specified number of groups (\code{n\_class} in the package) and result in a new symmetric matrix $D'$ with \code{n\_class} symbols---one for each of these groups. These symbols could be numbers or appropriate graphics.

Based on the Czekanowski's diagram, researchers should be able to identify clusters of observations, due to its inherent simplicity and compactness of presentation. 
For instance, this diagram was employed to discern clusters of counties in the Silesia Voivodeship based on their Internet availability \cite{KWar2015}.
However, this cluster identification has up to now been done manually, i.e., the researcher had to look at the diagram and decide where to draw the line between clusters of observations. In some cases this might be straightforward (e.g., Fig. 3 in \cite{KBarAVas2020}), but in others, even for small datasets, the placement of a border might be a challenging task. 
For example, if we go back to the internet availability data \cite{KWar2015}, which was also re--analyzed in \cite{KBarAVas2020}, we can clearly see distinct groupings of counties (Fig. $5$ in \cite{KBarAVas2020}). However, clear boundaries are not always easily distinguishable. The issue becomes even more visible when the data is generated by a mechanism that induces dependencies between observations.
We experimented \cite{KBarAVas2020} with generating phylogenetically structured data, and even when the seriation algorithm managed to recover the correct order, extracting the clades exactly by visual inspection of the diagram is not a feasible task (Fig. $1$ in \cite{KBarAVas2020}). 
Our aim in this work is to provide an algorithmic approach to identify clusters of points inside Czekanowski's diagram. 
The main idea behind, what we call \emph{Czekanowski's Clustering} is finding the partition points within the rearranged data to segment it into distinct clusters. 
We describe our approach in Alg. \ref{algExactCzClust} and visualize it in Fig. \ref{fig:flowchart}.


\section{Fuzzy C--means clustering}\label{secFuzzyCmeans}
Clustering techniques can be broadly classified into two categories: hard clustering and fuzzy clustering. 
Hard clustering partitions a dataset into distinct, non-overlapping clusters, assigning each data point to exactly one cluster. 
Hard clustering algorithms result in crisp, well-defined partitions of the data, with each data point having a binary membership status in a particular cluster. 
In contrast, fuzzy clustering is a technique that allows for the partial assignment of data points to different clusters. 
In fuzzy clustering, a data point's membership in a given cluster is determined by a membership function that assigns a real-valued weight to each point's degree of association with that cluster. 
This weight lies between $0$ and $1$, and indicates the degree to which a given point belongs to a particular cluster, enabling the incremental assessment of an observation's membership.
As a result, a data point can exist in multiple groups simultaneously, with varying degrees of membership. 
Such an approach enables a more nuanced understanding of data points' cluster memberships, accommodating points that are clearly within a cluster and those that are more ambiguously positioned. Such a property fits our setting very well---in Fig. \ref{fig:ExampleRMaCzek} the cars in the middle of the visible clusters will be certain members; but those on the edges of the clusters can be assigned both ways.

The Fuzzy C-Means algorithm (FCM) \cite{bezdek1984fcm, bezdek1981pattern}, proposed by James C. Bezdek in 1984, is a widely recognized fuzzy clustering algorithm. 
It runs on a predetermined number of clusters K and iteratively minimizes its objective function. 
\pkg{RMaCzek} uses the implementation provided by the \pkg{e1071} \proglang{R} package \cite{DMeyetal2022}, and we refer to Ch. $2.3$ in \cite{YLuo2022} for more details.
When designing the algorithm, we were faced with a decision: whether to apply a change point detection method directly to the Czekanowski’s Matrix, or to use the FCM before this step. After careful consideration and experimental validation, we chose the latter approach for several compelling reasons. Firstly, the computational cost associated with identifying change points is lower when using the membership matrix derived from FCM, as opposed to applying a change point detection method directly on the Czekanowski’s Matrix. The dimensionality of the membership matrix, determined by K, the number of specified clusters, is always lower than N, the number of samples in the dataset. This reduced dimensionality streamlines the change point detection process. Secondly, FCM enhances the separability of data, facilitating the subsequent change point detection method in more effectively discerning distinct clusters, especially in datasets characterized by complexity or overlapping features. The algorithm achieves this by computing degrees of membership, which inherently smoothens the data, thereby mitigating the impact of noise and outliers. This effectiveness is further supported by our experimental results, which demonstrate improved performance with the incorporation of FCM \cite{YLuo2022}. Lastly, compared to hard clustering methods, fuzzy clustering offers a more nuanced representation of cluster membership. This flexibility often makes initial data analysis more effective, helping to uncover underlying patterns that are useful for detecting changes in the data. Among various fuzzy clustering techniques, FCM is chosen for its reported quick convergence and its ability to perform comparably to maximum likelihood, especially when clusters are sufficiently separable \cite{SWieMKlo2015}. This algorithm also offers potential for future enhancements, such as the incorporation of different distance functions. It's important to note that while FCM itself is a clustering method, the use of a change point detection method in conjunction with FCM is crucial. This is because the samples in Czekanowski’s diagram are serialized, and the clustering revealed has to be contiguous. The change point detection method aids in identifying these contiguous clusters effectively, complementing the preliminary clustering done by FCM.

\section{E--divisive algorithm}\label{secEdivisive}
The E-divisive algorithm \cite{matteson2014nonparametric} is a hierarchical technique that is used to detect change points in series data. It recursively partitions a time series and uses a permutation test to determine change points. 
It can be used to analyze and understand trends, patterns, and anomalies in a wide range of applications.
The iterative estimation of change points involves identifying a new change point within an existing segment at each iteration by evaluating its statistical significance through a permutation test.
In particular, at iteration $k-1$, $k \geq 1$, we are given a current set of change points that have segmented the time series into $k$ segments, denoted as $S_1$, $S_2$, ..., $S_k$, and an estimated location for the next change point $\hat{\tau}_k$, with an associated test statistic value of $\hat{q}_k$.
We permute the observations within each of these segments, then re-estimate the position of the change point $\hat{\tau}_{k,r}$, and calculate the test statistic value, expressed as $\hat{q}_k^{(r)} $.
We repeat this permutation R times to find $\hat{q}_k^{(1)}, \hat{q}_k^{(2)}, \cdots, \hat{q}_k^{(R)}$.
An approximate p-value is subsequently calculated as $\hat{p} = \#\{ r:\hat{q}_{\kappa}^{(r)} \ge \hat{q}_{\kappa} \}/(R+1)$.
It is compared with a threshold p-value $p_0$, which is set to determine the statistical significance of the estimated change point. 
If the p-value $\hat{p}$ is found to be less than $p_0$, the current point is deemed significant, and the algorithm proceeds to estimate the next change point. 
Otherwise, the null hypothesis of the lack of existence of an additional change point cannot be rejected, and the algorithm terminates. 

\pkg{RMaCzek} uses the implementation provided by the \pkg{ecp} \proglang{R} package \cite{NJamDMat2014}.
A detailed description of the steps of the E--divisive algorithm can be found in Ch. $2.4$ of \cite{YLuo2022}.

\section{Czekanowski's clustering}
\begin{figure}[tbh!]
    \centering
Membership ma	\includegraphics[width=.9\linewidth]{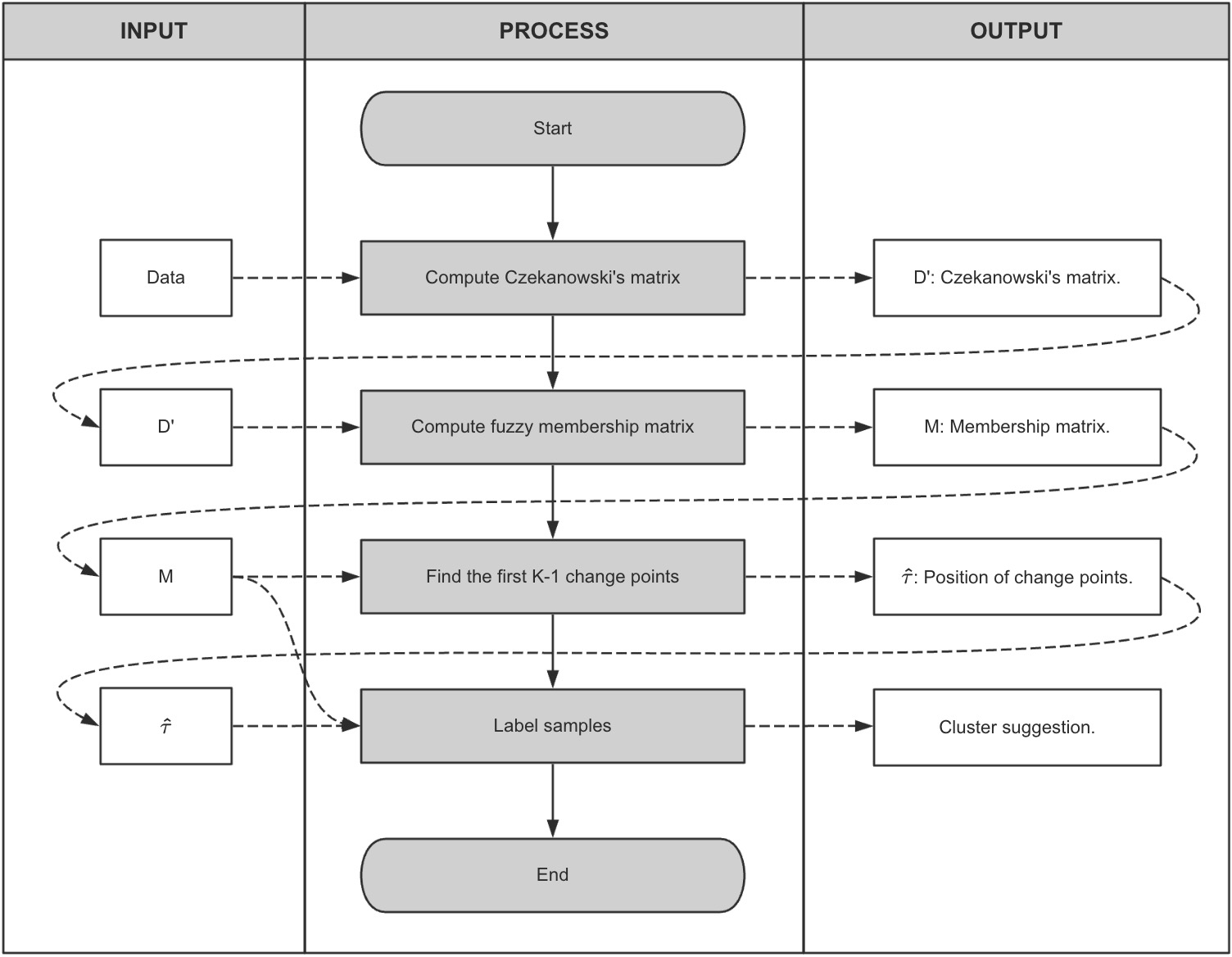}
	\caption{\label{fig:flowchart}Visualization of Czekanowski's Clustering Algorithm.} 
\vspace{-2ex}	
\end{figure}

In \cite{YLuo2022} an algorithm, that combines fuzzy C--means clustering, and the E--divisive algorithm, with the purpose of identifying clusters in Czekanowski's diagram was proposed. The membership matrix output by the fuzzy C--means clustering algorithm is passed as input to the E--divisive algorithm. Breakpoints are identified in the membership matrix, and these then define the clusters of observations. This method is presented, and subsequently visualize in Alg.~\ref{algExactCzClust} and Fig. \ref{fig:flowchart} ,and afterwords we describe the individual steps. A detailed
presentation of the multiple technical steps can be found in \cite{YLuo2022}. 

\begin{algorithm}[h]
\caption{Exact Czekanowski's Clustering Algorithm (see also Fig. \ref{fig:flowchart})
} \label{algExactCzClust}
\textbf{Input:}
\begin{itemize}\itemsep2pt \parskip0pt \parsep0pt
	\item[] $X$: The dataset.
	\item[] $K$: The pre--defined number of clusters.
\end{itemize}

\textbf{Output:}
\begin{itemize}\itemsep2pt \parskip1pt \parsep1pt
	\item[] $res\_cluster$: The clustering suggestion for each observation.
\end{itemize}


\begin{algorithmic}[1] 
    \State Create Czekanowski's diagram and associated matrix, $D'$.
    \State Obtain the membership matrix $M\in [0,1]^{K  \times N}$ 
    derived from a fuzzy clustering algorithm applied to $D'$.
	\State Find $K-1$ changing points of $M$ to get $K$ cluster intervals.
	\State Label observations of $D'$ in each cluster, resulting in $res\_cluster$. \\
	\Return $res\_cluster$
\end{algorithmic}
\end{algorithm}

\hfill \break
\textbf{STEP 1} \\
The creation of Czekanowski's diagram itself is well described in the literature and we direct the reader to, e.g., \cite{KBarAVas2020,JCze1909,JCze1913,YLuo2022,ASolPJas1999,AVas2019}.

\hfill \break
\textbf{STEP 2} \\
After creating Czekanowski's diagram, $D'$ we would like to identify groups of observations that are closely related---clusters of data. 
However, up to now the researcher had to manually read off the clusters from the diagram (as we would in, e.g., Fig.~\ref{fig:ExampleRMaCzek}). 
Even if clusters are obvious (though this should not always be expected), the exact point of delimitation might not be clear.
There can be great uncertainty with boundary observations. Hence, automatic clustering methods for Czekanowski's diagrams could greatly aid in this. 
A fuzzy C--means algorithm is applied to the matrix $D'$ to divide it into $K$ clusters.
As a result we obtain a membership matrix $M\in [0,1]^{K \times N}$, where each observation has a value indicating the likeliness of it belonging to each of the $K$ clusters. \\
\hfill \break
\textbf{STEP 3} \\
In contrast to general fuzzy clustering methods,
we do not assign observations to clusters according to
the highest membership values. This would result in a potential, loss of arrangement of the observations by the
seriation algorithm. 
Instead, we determine the clustering, by locating $K-1$ segmentation points of the membership matrix $M$---splitting it into $K$ intervals, each interval of observations is a cluster. In Fig.~\ref{fig:membership} 
we present a membership matrix showing that serialized values exhibit significant numerical shifts, which typically represent the breakpoints between clusters.

The breakpoints in the membership matrix $M$ are identified using the previously presented E--divisive algorithm.
As a result, we obtain $K$ groups of observations---the clustering of our seriated data. \\
\hfill \break
\textbf{STEP 4} \\
After the change points have been identified we can assign to each element a label---the identifier of the cluster it belongs to. 
We can then display the inferred clusters on the Czekanowski's diagram corresponding to $D'$, e.g., as background shading, as in 
Fig.~\ref{fig:OLOWBCCzkDiag}.
The final step is for the researcher to connect the
clusters with domain knowledge.

\section{\label{sec:examples}Example analysis: Wisconsin Breast Cancer data set.} Breast Cancer Wisconsin (Original) \cite{WBC} is a data set collected by Dr. William H. Wolberg from his clinical cases between 1989 and 1991. 
\begin{table}[tbh!]
\footnotesize
\caption{\label{tab:WBCdata}Summary of the Wisconsin Breast Cancer dataset.}

\resizebox{\textwidth}{!}{%
\renewcommand{\arraystretch}{1.3}
\begin{tabular}{!{\vrule width 1pt}c!{\vrule width 1pt}c!{\vrule width 1pt}c!{\vrule width 1pt}}
\ChangeRT{1pt}\multirow{11}{*}{Atrributes}        & {\multirow{2}{*}{Types}} & \multirow{2}{*}{Value Range} \\
                      & {}                            &                 \\ \ChangeRT{1pt}
                      & {Clump Thickness}             & 1 - 10          \\
                      & {Uniformity of Cell Size}     & 1 - 10          \\
                      & {Uniformity of Cell Shape}    & 1 - 10          \\
                      & {Marginal Adhesion}           & 1 - 10          \\
                      & {Single Epithelial Cell Size} & 1 - 10          \\
                      & {Bare Nuclei}                 & 1 - 10          \\
                      & {Bland Chromatin}             & 1 - 10          \\
                      & {Normal Nucleoli}             & 1 - 10          \\
                      & {Mitoses}                     & 1 - 10          \\ \ChangeRT{1pt}
\multirow{2}{*}{Class Distribution} & {2 = Benign}             & 4 = Malignant                \\ \cline{2-3}
                      & {458   $(65.5522\%)$}           & 241 $(34.4778\%)$ \\ \ChangeRT{1pt}
Number of   Instances & \multicolumn{2}{c!{\vrule width 1pt}}{699}                                           \\ \ChangeRT{1pt}
Number of   Attributes              & \multicolumn{2}{c!{\vrule width 1pt}}{11 (ID, diagnosis, 9 real-valued input features)}      \\ \ChangeRT{1pt}
Missing values        & \multicolumn{2}{c!{\vrule width 1pt}}{16}                                            \\ \ChangeRT{1pt}
\end{tabular}%
}\\[-2ex]
\end{table}

It comprises $699$ measurements from women sampled at eight time points during this period.
Each observation includes nine characteristics and a diagnosis (either benign or malignant) as its class label.
As the Bare Nuclei feature has 16 missing values, we only consider fully observed patients $683$.
In Tab.~\ref{tab:WBCdata}, we provide a summary of the data set.

We investigated if \pkg{RMaCzek} can identify whether a tumour is benign or malignant, and present the data in an informative manner.
We experiment with five seriation methods from the \pkg{seriation} package \cite{MHahKHorCBuc2008}---OLO (\cite{ZBaretal2001}, in two versions), HC, GW (both versions of hierarchical clustering, for GW see \cite{GGruHWai1972}), and SPIN\_NH (optimization of an appropriate matrix functional, see \cite{DTsaetal2005}).
OLO (Optimal Leaf Order) with average linkage is the default method in \pkg{RMaCzek}, as it has been found very effective in previous studies \cite{KBarAVas2020,YLuo2022,AVas2019}.
In fact, in \cite{YLuo2022}, $1000$ synthetic datasets using Gaussian mixture models were generated, and the performance of $32$ serialization methods on these datasets was evaluated in order to find effective serialization methods. The top five methods were those considered in this work. Here, we also tested OLO with Ward's linkage. OLO minimizes the Hamiltonian path length of a tree, where the leaves are the observations, derived from the distance matrix between the observations. Linkage is the method of calculating the distance between two clusters when hierarchically constructing the tree\footnote{\url{https://www.mathworks.com/help/stats/linkage.html}}—\emph{average} is the average distance between all pairs of objects in the true clusters; \emph{Ward's} method calculates the increase in the within-cluster sum of squares (sum of squares of distances of the objects to the cluster's centroid) after the joining of the two clusters.
A very recent discussion concerning, among others, the OLO method, can be found in \cite{DAliCZir2023}.
The two other considered hierarchical clustering methods---HC and GW---are also used with Ward's linkage.

\begin{figure}[tbh!]
	\includegraphics[width=\linewidth]{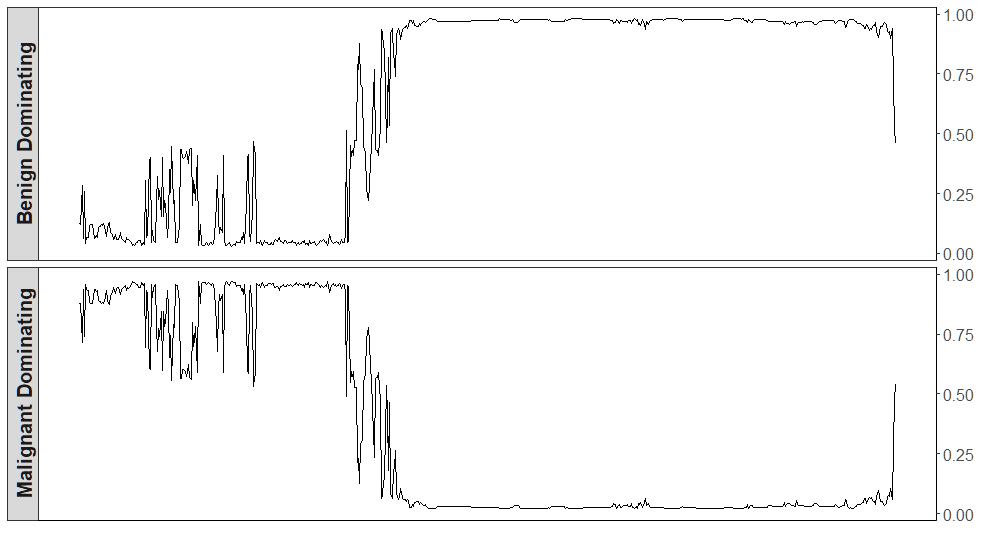}
	\caption{Membership Matrix ($K=2,~N=683$) for the WBC dataset derived from the $D'$ matrix obtained under the OLO\_average seriation method.}
	\label{fig:membership}
\end{figure}

We used a number of common statistics to evaluate the methods---accuracy, recall, precision, F1, and Cohen's $\kappa$. 
They are based on counting how many items are correctly classified in a given cluster, i.e., the true positive (TP), true negative (TN), false positive (FP), and false negative (FN) counts. 
We assumed $K=2$, and hence we can compare our clustering results with the true labels (benign/malignant) by permuting every possible labelling of the clusters and then calculating the corresponding accuracy for each permutation. 
We then take the assignment of labels to clusters with the highest accuracy for assessing the algorithm's performance. 
We can see, in Fig.~\ref{fig:membership}, that this is a majority rule labelling (in our $K=2$ case), except we describe it for the general $K$ situation. 
The choice of positive and negative is arbitrary in our case, in fact we consider both situations when benign and malignant classes are considered ``positives'' to see how the seriation methods perform. The scores are defined as follows
\[
\begin{array}{ll}
accuracy = \frac{TP+TN}{TP+TN+FP+FN}, & recall = \frac{TP}{TP+FN}, \\~\\
precision = \frac{TP}{TP+FP}, & F1 = \frac{2 \cdot precision \cdot recall}{precision+recall} = \frac{2 \cdot TP}{2 \cdot TP+FP+FN},
\end{array}
\]
and Cohen's $\kappa$ measures the agreement between two raters (i.e., the benign ``rater'' rating as benign and not--benign, and similarly the malignant ``rater''),
\begin{align*}
\kappa &= \frac{n_{bb}n_{mb}+n_{bm}n_{mm}}{N^{2}}\\
&={  
\frac{2(TP\cdot TN-FN \cdot FP)}{(TP+FP)\cdot(FP+TN)+(TP+FN)\cdot(FN+TN)}},
\end{align*}
were $n_{ij}$ is how many observations rater $i$ (benign/malignant) classified as $j$ (benign/malignant) and $N$ is the total number of observations.

Furthermore, we also evaluated the performance of the methods using two additional metrics: the path length and the factor $U_m$.
The factor $U_m$, introduced by Sotysiak and Jaskulski \cite{ASolPJas1999}, serves as a measure of matrix arrangement, used in the calculation of the optimization function.
It represents the average ratio of the squared distance of a cell from the diagonal to the value of the cell, incremented by one to prevent division by zero. The $U_m$ factor is employed to compare matrices, in some cases, a lower value indicates a more optimal arrangement:
$$
U_m=\frac{2}{n^2}\sum_{j=1}^n\sum_{i=j+1}^n\frac{(i-j)^2}{W_{\pi(i),\pi(j)}+1},
$$
where $W_{ij}$ denotes the distance between objects at positions $i$ and $j$, and $\pi$ is a permutation.

The OLO seriation method, as previously described, functions by minimizing the total distance of the path connecting the leaves in a specified order, consistent with a tree structure.
We consider each observation point as a leaf, with the distance between observations represented by the length of the path linking the pair
\cite{ZBaretal2001},
$$
L_{path\_length} = \sum_{i=1}^{n-1}W_{\pi(i),\pi(i+1)}.
$$

In Fig.~\ref{fig:membership} we can see how the membership values to each cluster (step $2$ in \
Fig.~\ref{fig:flowchart}) change along the seriated observations.

In this study, specific parameters were selected for each algorithm to optimize the analysis according to our data set and research goals. For a comprehensive understanding of these parameters, we refer the reader to \pkg{RMaCzek}'s manual, \cite{DMeyetal2022}, and \cite{NJamDMat2014} for detailed explanations. \\

\begin{table}[t]
  \footnotesize
  \caption{ \label{tab:algorithm-parameters} Parameters used in the algorithms for the experiment.}
  
  \centering
  \renewcommand{\arraystretch}{1.3}

  \begin{tabular}{!{\vrule width 1pt}>{\centering\arraybackslash}m{3.5cm}|>{\centering\arraybackslash}m{1.5cm}|>{\centering\arraybackslash}m{7cm}!{\vrule width 1pt}}
    \ChangeRT{1pt}

    \textbf{Parameter} & \textbf{Value} & \textbf{Description} \\ 
    \ChangeRT{1pt}

    \multicolumn{3}{!{\vrule width 1pt}c!{\vrule width 1pt}}{\textbf{Fuzzy C-Means (FCM)}} \\ 
    \ChangeRT{1pt}
    Centers & 2 & Representing Benign and Malignant. \\ \hline
    Max Iterations & 100 & Maximum number of iterations. \\ \hline
    Distance Function & Euclidean & $d(x,y) = \sqrt{\sum_i (x_i - y_i)^2}$. \\ \hline
    Fuzzification Degree & 2 & $m$ in the objective function \( J_m \), where \( J_m = \sum_{i=1}^{N} \sum_{j=1}^{C} u_{ij}^m \cdot d(x_i, c_j)^2 \). Here, \( C \) is the number of clusters, \( u_{ij} \) is the membership degree of data point \( i \) in cluster \( j \), and \( c_j \) is the centroid of cluster \( j \). \\ \hline

    \ChangeRT{1pt}
    \multicolumn{3}{!{\vrule width 1pt}c!{\vrule width 1pt}}{\textbf{E–divisive}} \\ 
    \ChangeRT{1pt}
    Significance Level & 0.05 & Statistical significance level. \\ \hline
    Max Permutations & 199 & Number of permutations R in the E--divisive algorithm. Refer to Section \ref{secEdivisive} for details. \\ \hline
    Change Point Estimates & 1 & Number of change points to estimate. \\ \hline
    Min Observations & 2 & Minimum observations between change points. \\ \hline
    Moment Index & 1 & Determines distance in/between segments. \\ 

    \ChangeRT{1pt}
    \multicolumn{3}{!{\vrule width 1pt}c!{\vrule width 1pt}}{\textbf{Czekanowski’s Diagram}} \\ 
    \ChangeRT{1pt}
    n\_classes & 5 & Number of classes. \\ \hline
    Distance Function & Euclidean & $d(x,y) = \sqrt{\sum_i (x_i - y_i)^2}$. \\ \hline
    Data Scaling & True & z--score: \( \hat{x} = \frac{x - \overline{x}}{s} \). \\

    \ChangeRT{1pt}
  \end{tabular}
\end{table}
\begin{figure}[h!tb]
	\includegraphics[width=\linewidth]{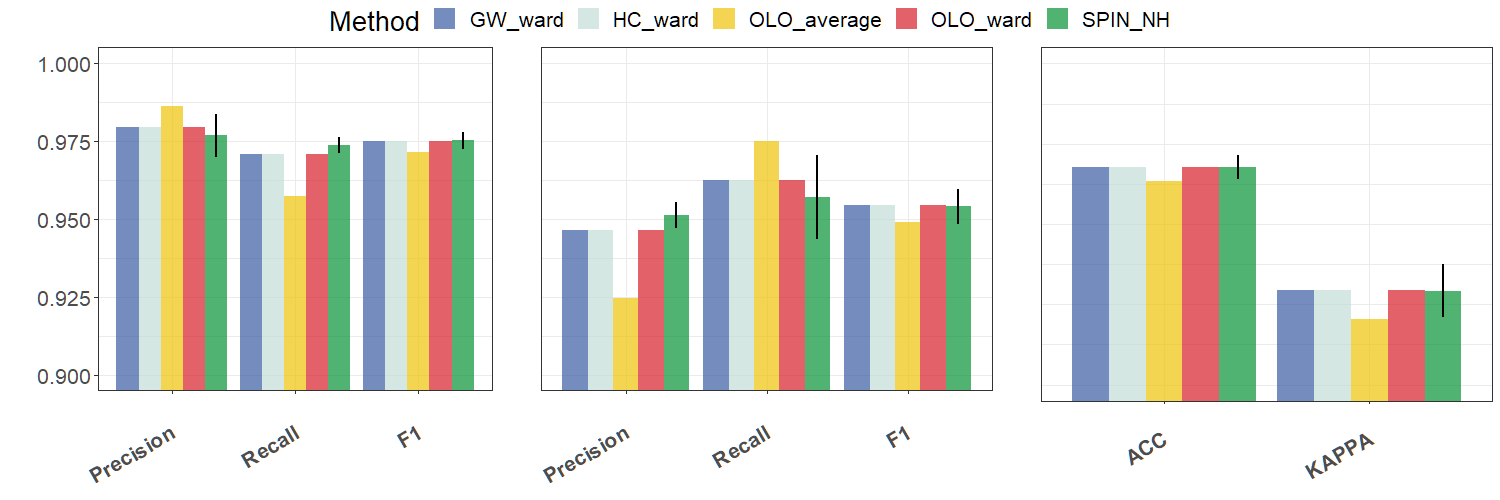}
	\caption{Evaluation of seriation methods on the WBC dataset. Left: Precision, Recall, and F1-score with Benign as the positive label. Centre: Precision, Recall, and F1-score with Malignant as the positive label. Right: Overall Accuracy and Cohen's $\kappa$ score.}
	\label{fig:evalWBC}
\end{figure}

In Fig.~\ref{fig:evalWBC} and Tab.~\ref{tabPerfWBC}, it is evident that all the evaluated methods exhibit similar performance characteristics.
However, a notable observation from the precision and recall statistics is that the OLO\_average method tends to slightly misclassify benign tumours as malignant.

\begin{table}[tbh!]
\footnotesize
\caption{\label{tabPerfWBC}Performance of \pkg{RMaCzek}'s clustering on Czekanowski's diagram, by different seriation methods, of the WBC dataset. Results for SPIN\_NH are presented as mean $\pm$ standard deviation.} 

\resizebox{\textwidth}{!}{%
\renewcommand{\arraystretch}{1.3}
\begin{tabular}{!{\vrule width 1pt}c!{\vrule width 1pt}c|c|c|c|c!{\vrule width 1pt}}
\ChangeRT{1pt}
{}                  & \multicolumn{1}{c|}{\textbf{GW\_ward}} & \multicolumn{1}{c|}{\textbf{HC\_ward}} & \multicolumn{1}{c|}{\textbf{OLO\_average}} & \multicolumn{1}{c|}{\textbf{OLO\_ward}} & \textbf{SPIN\_NH }       \\
\ChangeRT{1pt}
{\textbf{Accuracy}}          & \multicolumn{1}{c|}{0.9678}   & \multicolumn{1}{c|}{0.9678}   & \multicolumn{1}{c|}{0.9634}       & \multicolumn{1}{c|}{0.9678}    & 0.9678$\pm$0.003595   \\
{\textbf{Kappa}}             & \multicolumn{1}{c|}{0.9295}   & \multicolumn{1}{c|}{0.9295}   & \multicolumn{1}{c|}{0.9205}       & \multicolumn{1}{c|}{0.9295}    & 0.9292$\pm$0.008244   \\
{\textbf{Path Length}}       & \multicolumn{1}{c|}{611.4090} & \multicolumn{1}{c|}{686.3781} & \multicolumn{1}{c|}{592.5447}     & \multicolumn{1}{c|}{587.3688}  & 975.4746$\pm$38.77697 \\
{\textbf{$U_m$ Factor}}      & \multicolumn{1}{c|}{15352.73} & \multicolumn{1}{c|}{17318.13} & \multicolumn{1}{c|}{15821.93}     & \multicolumn{1}{c|}{16038.55}  & 15077.04$\pm$138.1793 \\
\ChangeRT{1pt}
\multicolumn{6}{!{\vrule width 1pt}c!{\vrule width 1pt}}{\textbf{Benign (as positive)}}                                                                                                                                                                   \\
\ChangeRT{1pt}
{\textbf{Precision}}         & \multicolumn{1}{c|}{0.9795}   & \multicolumn{1}{c|}{0.9795}   & \multicolumn{1}{c|}{0.9861}       & \multicolumn{1}{c|}{0.9795}    & 0.9768$\pm$0.006828 \\
{\textbf{Recall}}            & \multicolumn{1}{c|}{0.9707}   & \multicolumn{1}{c|}{0.9707}   & \multicolumn{1}{c|}{0.9572}       & \multicolumn{1}{c|}{0.9707}    & 0.9736$\pm$0.002545 \\
{\textbf{F1-score}}          & \multicolumn{1}{c|}{0.9751}   & \multicolumn{1}{c|}{0.9751}   & \multicolumn{1}{c|}{0.9714}       & \multicolumn{1}{c|}{0.9751}    & 0.9752$\pm$0.002632 \\
\ChangeRT{1pt}
\multicolumn{6}{!{\vrule width 1pt}c!{\vrule width 1pt}}{\textbf{Malignant (as positive)}}                                                                                                                                                                \\
\ChangeRT{1pt}
{\textbf{Precision}}         & \multicolumn{1}{c|}{0.9465}   & \multicolumn{1}{c|}{0.9465}   & \multicolumn{1}{c|}{0.9246}       & \multicolumn{1}{c|}{0.9465}    & 0.9513$\pm$0.004185 \\
{\textbf{Recall}}            & \multicolumn{1}{c|}{0.9623}   & \multicolumn{1}{c|}{0.9623}   & \multicolumn{1}{c|}{0.9749}       & \multicolumn{1}{c|}{0.9623}    & 0.9569$\pm$0.013450 \\
{\textbf{F1-score}}          & \multicolumn{1}{c|}{0.9544}   & \multicolumn{1}{c|}{0.9544}   & \multicolumn{1}{c|}{0.9491}       & \multicolumn{1}{c|}{0.9544}    & 0.9540$\pm$0.005631  \\
\ChangeRT{1pt}
\end{tabular}%
}
\end{table}

\begin{table}[tbh!]
\caption{\label{tabconfusion}Confusion matrix of \pkg{RMaCzek}'s clustering on Czekanowski's diagram, by different seriation methods, of the WBC data set. B for Benign and M for Malignant. Results for SPIN\_NH are presented as mean $\pm$ standard deviation.}

  \resizebox{\textwidth}{!}{%
    \renewcommand{\arraystretch}{1.3}
    \begin{tabular}{!{\vrule width 1pt}cc!{\vrule width 1pt}c|c!{\vrule width 1pt}c|c!{\vrule width 1pt}c|c!{\vrule width 1pt}c|c!{\vrule width 1pt}c|c!{\vrule width 1pt}}
      \ChangeRT{1pt}
      {} & {} & \multicolumn{10}{c!{\vrule width 1pt}}{\textbf{ACTUAL}} \\ 
      \Xcline{3-12}{1pt}
      &  &
      \multicolumn{2}{c!{\vrule width 1pt}}{\textbf{GW\_ward}} &
      \multicolumn{2}{c!{\vrule width 1pt}}{\textbf{HC\_ward}} &
      \multicolumn{2}{c!{\vrule width 1pt}}{\textbf{OLO\_average}} &
      \multicolumn{2}{c!{\vrule width 1pt}}{\textbf{GW\_ward}} &
      \multicolumn{2}{c!{\vrule width 1pt}}{\textbf{SPIN\_NH}} \\ \cline{3-12}
      &  &
      \textbf{B} & \textbf{M} & \textbf{B} & \textbf{M} & \textbf{B} & \textbf{M} & \textbf{B} & \textbf{M} & \textbf{B} & \textbf{M} \\
      \ChangeRT{1pt}
      \multirow{2}{*}{\textbf{RESULT}}  &
      \multicolumn{1}{!{\vrule width 1pt}c!{\vrule width 1pt}}{\textbf{B}} & 431 & 9 & 431 & 9 & 425 & 6 & 431 & 9 & 432.28$\pm$1.13 & 10.30$\pm$3.21 \\
      \cline{2-12}
      {} & \multicolumn{1}{!{\vrule width 1pt}c!{\vrule width 1pt}}{\textbf{M}} & 13 & 230 & 13 & 230 & 19 & 233 & 13 & 230 & 11.72$\pm$1.13 & 228.70$\pm$3.21 \\
      \ChangeRT{1pt}
    \end{tabular}%
  }\\[2ex]
\end{table}

\begin{figure}[bh!]
	\centering
        \subfloat{{\includegraphics[width=.3\textwidth]{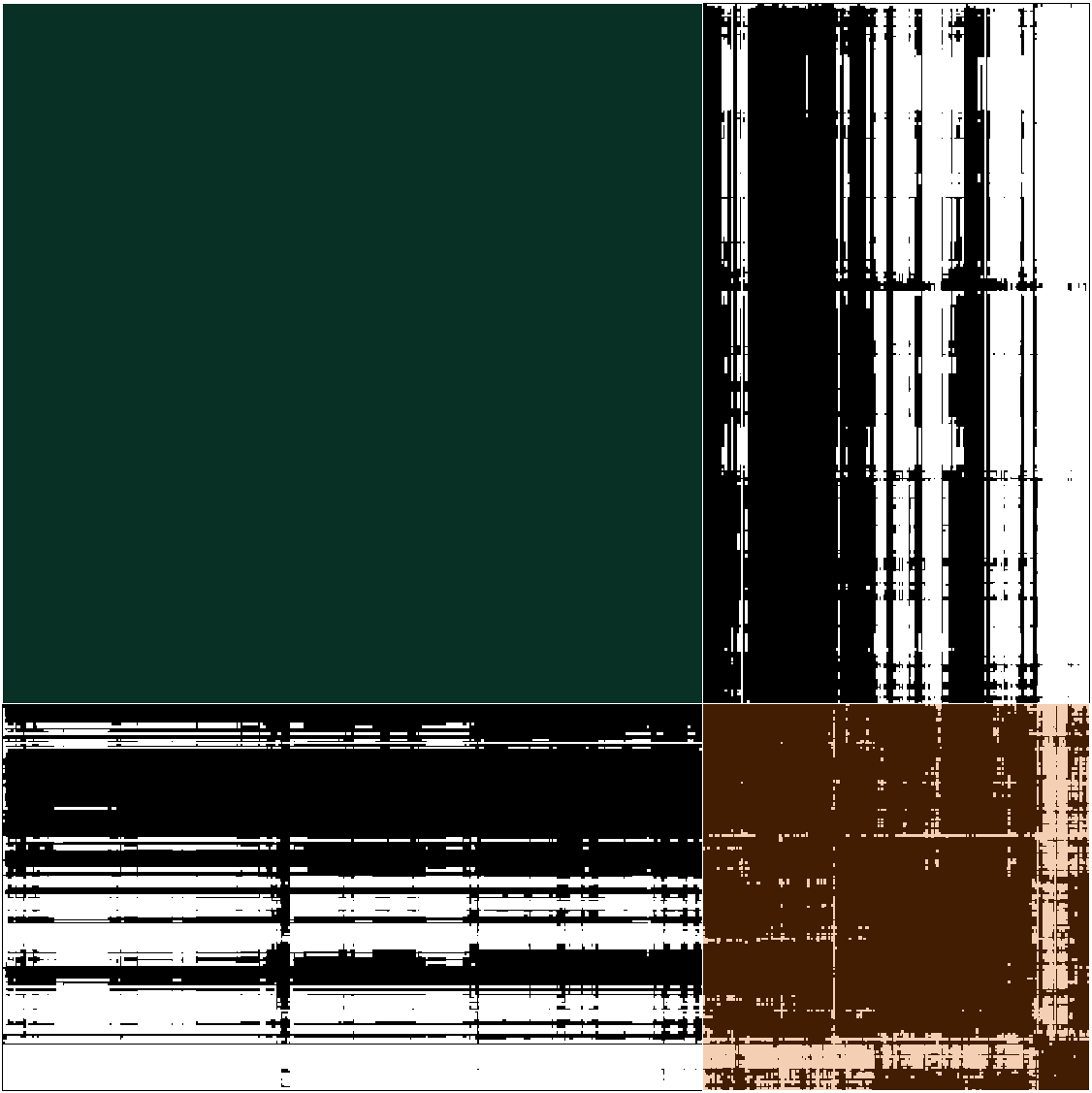}}} \hspace{1em}
    	\subfloat{{\includegraphics[width=.3\textwidth]{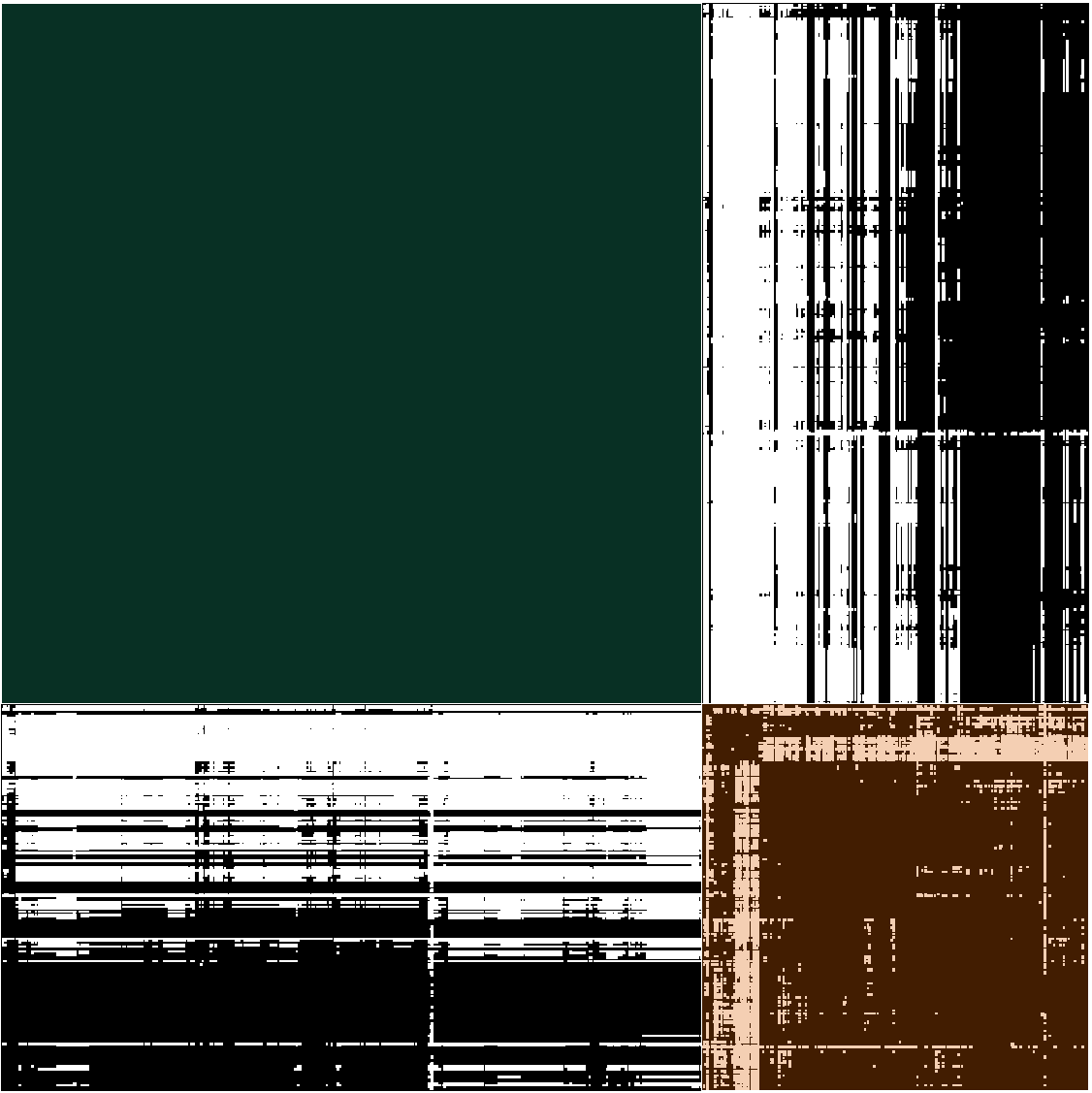}}} \hspace{1em}
    	\subfloat{{\includegraphics[width=.3\textwidth]{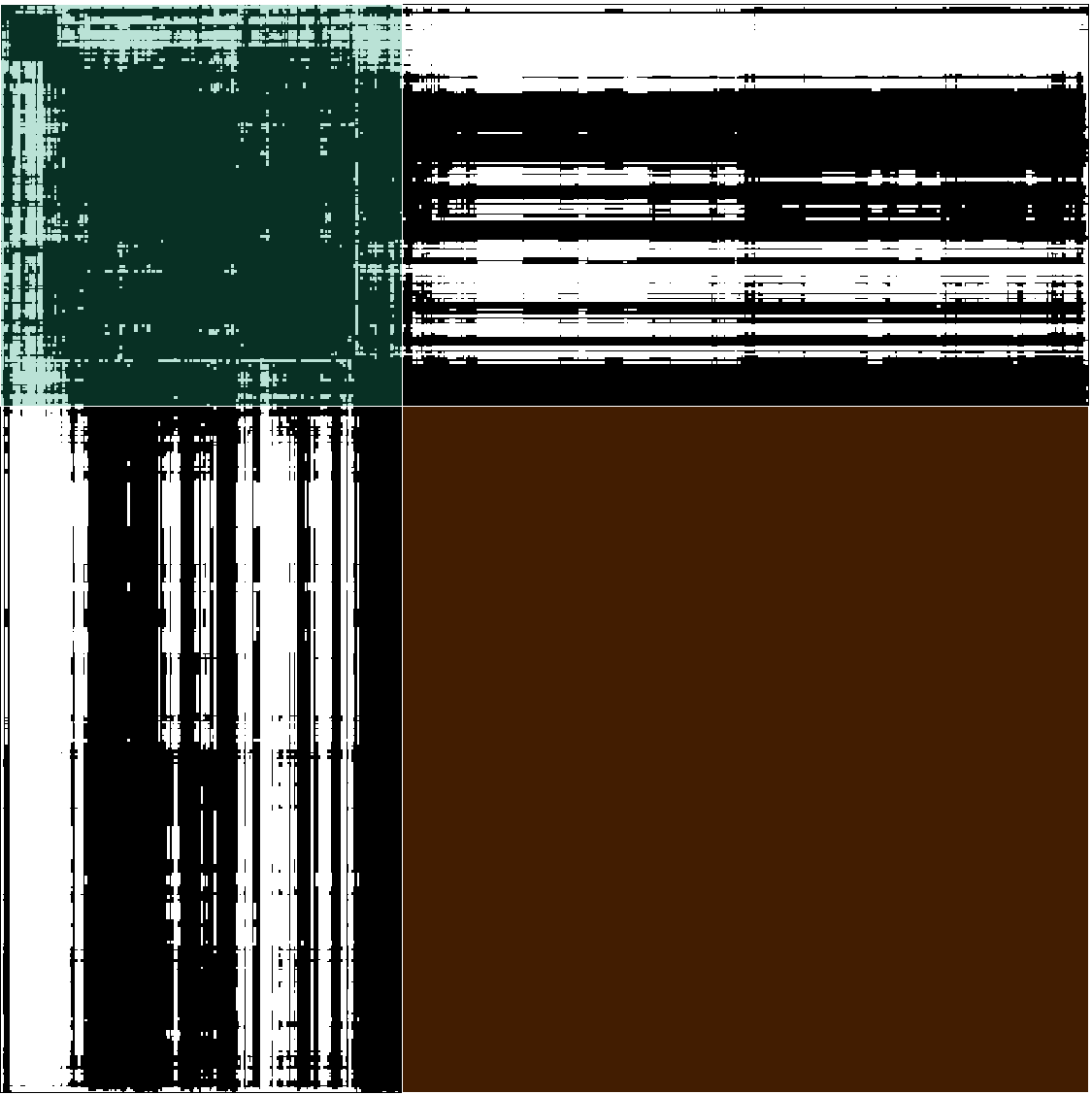}}}
     
    	\subfloat{{\includegraphics[width=.3\textwidth]{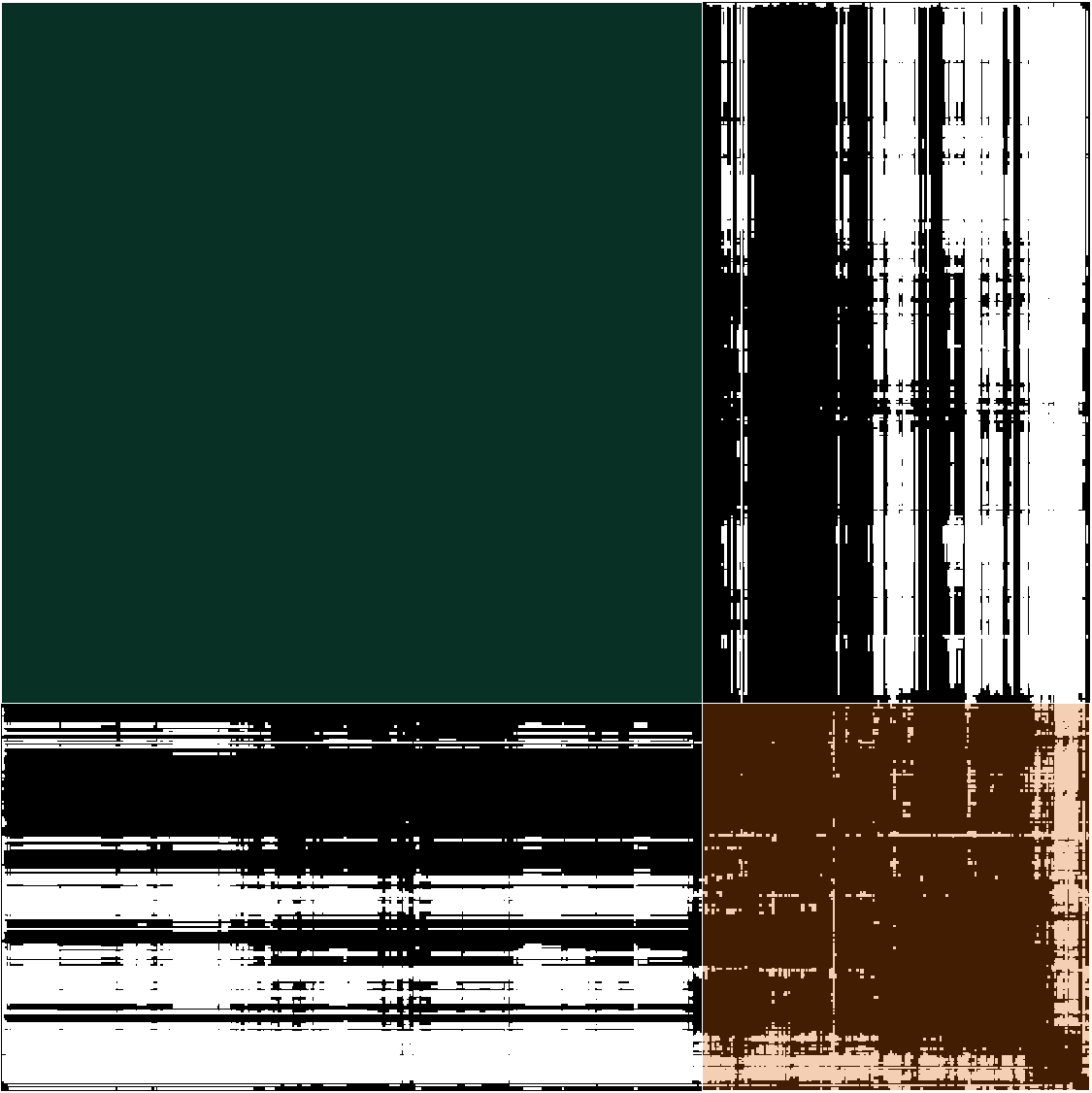}}} \hspace{1em}
    	\subfloat{{\includegraphics[width=.3\textwidth]{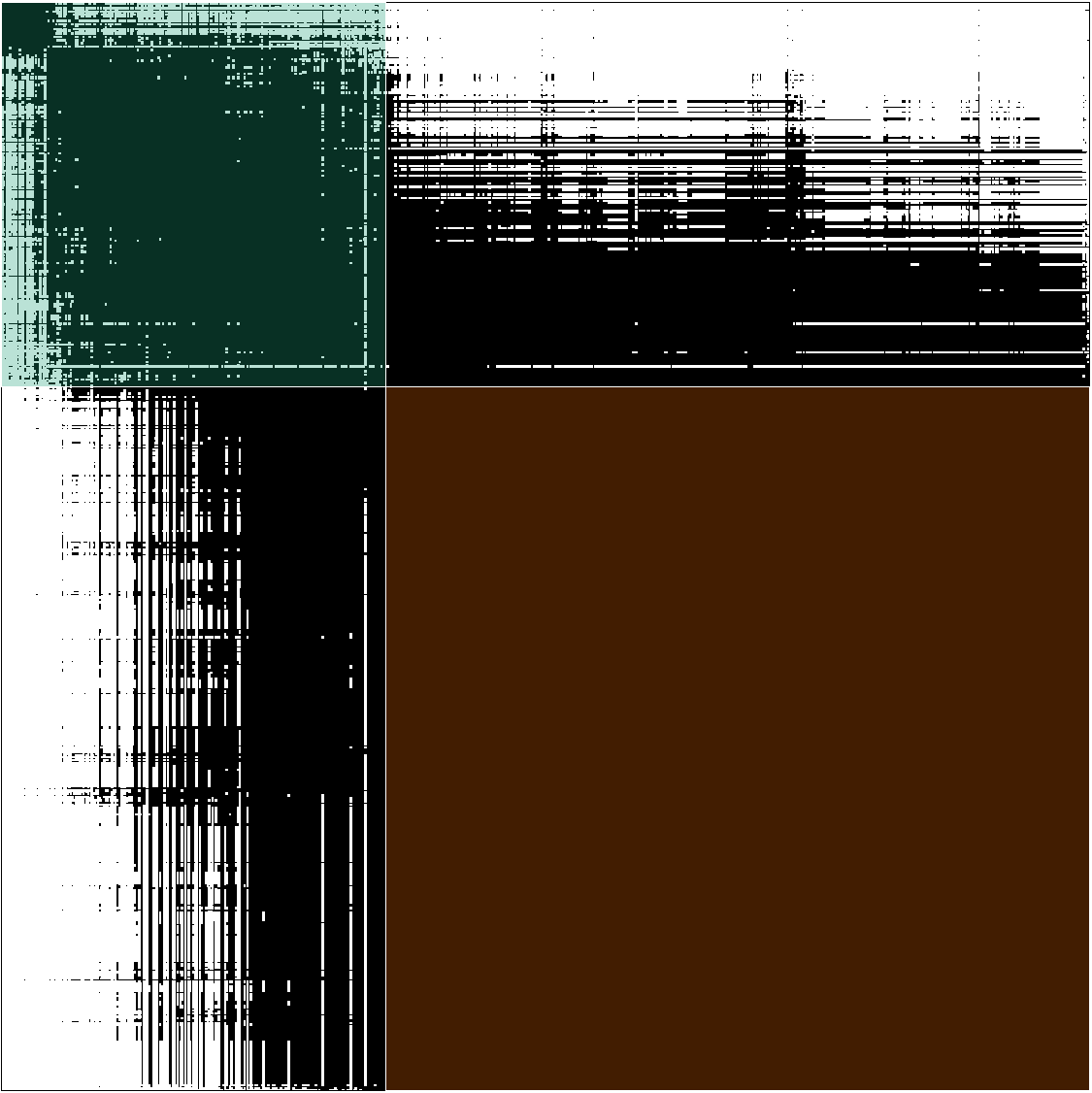}}}
	\caption{\label{fig:OLOWBCCzkDiag}Czekanowski’s diagrams based on  serialization with GW\_ward, HC\_ward, OLO\_average, OLO\_ward and SPIN\_NH methods (the order of the methods here corresponds to the order in the graphs in the figure).The shaded regions are the two clusters---the larger one is the benign dominating, while the smaller one, the malignant dominating. We can see that seriation with the HC\_ward gives a clear boundary between the clusters, while, on the other hand, e.g., SPIN\_NH does not provide a significant clustering tendency, and in OLO\_average's case the found clustering could deviate from visual discrimination.
 }
\end{figure}

\section{Conclusions.} In this study, we have enhanced the \pkg{RMaCzek} \proglang{R} package by integrating the cluster identification within Czekanowski's diagram.  
This advancement offers researchers an objective tool for cluster creation, moving beyond solely manual and subjective approaches.
What makes our work distinct from other clustering approaches is the philosophy---often one first constructs the clusters, based on the original variables, and then only displays the similarity between datapoints, e.g., through a heatmap. We do it the other way round, first rearrange the data points in one dimension; visualize the similarity matrix; and then indentify clusters inside this matrix.
We have illustrated the possibilities of our approach on a classical breast cancer data set. 
Our observations indicate that while top-serialization methods show similar performance, OLO with average linkage tends to slightly increase misclassification rates.
Future work should include a detailed examination of the combination of fuzzy clustering methods and Czekanowski's diagram, especially since the diagram inherently suggests soft boundaries between clusters.

Our proposed method is a valuable addition to the toolkit of applied researchers, potentially uncovering features that might be overlooked by other methods. As illustrated in Fig. \ref{fig:OLOWBCCzkDiag}, our approach not only distinctly visualizes clusters of different tumor types, but also highlights more nuanced aspects. Interestingly, Czekanowski's diagram reveals that some malignant tumors, located at the core of the malignant cluster, exhibit similarities to a range of samples, including benign ones. This observation suggests the need for a follow-up investigation into samples that, while classified within a specific cluster, still show a high degree of similarity to samples from other clusters.

\section{Software Availability}
\pkg{RMaCzek} can be found at \url{https://cran.r-project.org/web/packages/RMaCzek/}
and \url{https://github.com/krzbar/RMaCzek/}. 
The data and \proglang{R} scripts that allow the replication of the work are available at \url{https://github.com/krzbar/RMaCzek_KKZMBM2023}. 

\vspace{6pt} 
\authorcontributions{K.B. conceived the ideas and designed the methodology; Y.L. developed and implemented the algorithms, analyzed the data, and ran all the numerical experiments, all under K.B.'s mentorship; K.B. and Y.L. jointly contributed to the writing. Author ordering is alphabetical.
}
\funding{KB's research is supported by an ELLIIT Call C grant.
 Polish Mathematical Society remains neutral with respect to jurisdiction claims in published maps and institutional affiliations.}

\conflictsofinterest{The authors declare no conflict of interest.
}     
\acknowledgments{We would like to thank two anonymous Reviewers whose comments have improved our work. This work is based on YL’s master thesis in \textsl{Statistics and Machine Learning} “\textit{Czekanowski’s Clustering: Development of Visualization Possibilities of the RMaCzek Package} ' (2022) written in the \textbf{ Division of Statistics and Machine Learning, Department of
Computer and Information Science, Linköping University} (v.~\citeauthor{YLuo2022}~(\citeyear{YLuo2022})).}    


{\begin{center}\large\textbf{References}\end{center}}


\bigskip
\setcounter{section}{0}
\selectlanguage{polish}
\Polskitrue
\subjclass[2010]{62H99; 62-04; 92B10} 
\keywords{diagram czekanowskiego, kraniometria, metody odleg{\l }o{\'s}ci wielocechowych, rozw{\'o}j ludzko{\'s}ci}
%

\begin{center}
{\bf Identyfikacja skupień w diagramie Czekanowskiego.}\\ 
\href{\repo/7259}{Krzysztof Bartoszek i Ying Luo}
\end{center}
\medskip

\begin{abstract}{Diagram Czekanowskiego ma na celu zaprezentowanie podobie{\'n}stw wewn{\k a}trz pr\'obki statystycznej. 
Najcz{\k e}{\'s}ciej wida{\'c} na nim wyra{\'z}ne grupowania 
element{\'o}w. Jednak{\.z}e dok{\l }adne wyznaczenie granic mi{\k e}dzy skupieniami nie jest trywialnym zagdnieniem.
W niniejszej pracy przedstawiamy rozszerzon{\k a} wersj{\k e}
pakietu \pkg{RMaCzek}, kt{\'o}ra pozwala na analiz{\k e} skupie{\'n} w diagramach Czekanowskiego.}
\end{abstract}

\selectlanguage{polish}
\Polskitrue
\vspace{-3ex}
\begin{minipage}[b]{\linewidth}\small
\begin{wrapfigure}{l}{2.6cm}
  \vspace{-10pt}
  \begin{center}
    \includegraphics[width=2.6cm]{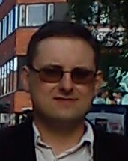}
  \end{center}
  \vspace{-10pt}
\end{wrapfigure}
\noindent \emph{Krzysztof Bartoszek}\footnote{\selectlanguage{english}References to his research papers are found in MathSciNet under  \href{http://www.ams.org/mathscinet/search/author.html?mrauthid= 793554}{ID: 793554} and the European Mathematical Society, FIZ Karlsruhe, and the Heidelberg Academy of Sciences bibliography database known as zbMath under \href{https://zbmath.org/authors/?q=ai: 	bartoszek.krzysztof}{ai:Bartoszek.Krzysztof}.} (ur. $1984$ w Bydgoszczy), obecnie wykładowca statystyki na Uniwersytecie w Link\"opingu. Absolwent informatyki Politechniki Gdańskiej (mgr inż.), biologii obliczeniowej Uniwersytetu w Cambridge (MPhil). Doktorat z statystyki matematycznej pod kierunkiem Serika Sagitova uzyskał w $2013$ r. na Uniwersytecie w G\"oteborgu. Jego główne zainteresowania związane są z procesami stochastycznymi w filogenetyce.
\hspace{2cm} 
\end{minipage}

\begin{minipage}[b]{\linewidth}\small
\begin{wrapfigure}{l}{2.6cm}
  \vspace{-10pt}
  \begin{center}
    \includegraphics[width=2.6cm]{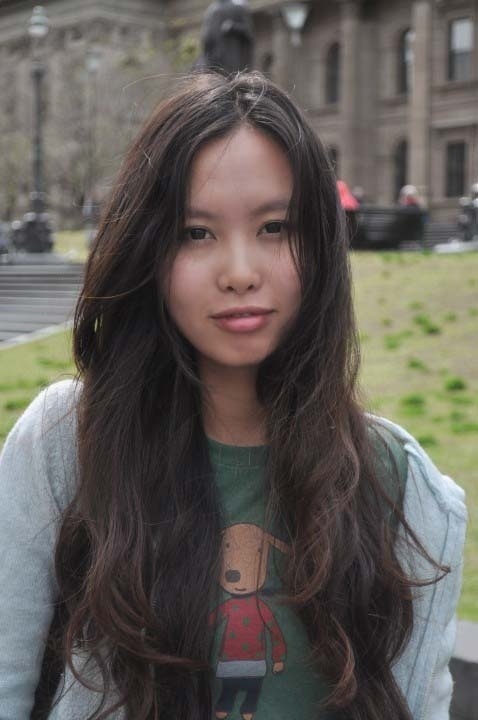}
  \end{center}
  \vspace{-10pt}
\end{wrapfigure}
\noindent \emph{Ying Luo}\footnote{\selectlanguage{english}References to her research papers can be found in \href{https://orcid.org/0009-0002-1298-0037}{ORCID}.
} (urodzona w 1986 roku w Xinjiang, Chiny) obecnie pracuje nad doktoratem z bioinformatyki na Uniwersytecie
w Linköpingu. Absolwentka studiów magisterskich z informatyki
na Królewskim Instytucie Technologii w Melbourne (2012) oraz
uczenia maszynowego na Uniwersytecie w Linköpingu (2022). Jej
akademicka droga charakteryzuje się trwałą pasją do badania
własności algorytmów i nieustanną chęcią wyłuskiwania wiedzy
z dostępnych danych. Rozwija teraz narzędzia, które wykorzystując potęgę uczenia maszynowego i modeli matematycznych, w
oparciu o dane empiryczne, będą służyć do analizy dużych struktur molekularnych.
\hspace{2cm} 
\end{minipage}
\vspace{1ex}
\label{koniec}
{\Koniec}
\end{document}